# MASER: A Science Ready Toolbox for Low Frequency Radio Astronomy

**Baptiste Cecconi**[1,2], **Alan Loh**[1], **Pierre Le Sidaner**[3], **Renaud Savalle**[3], **Xavier Bonnin**[1], **Quynh Nhu Nguyen**[1], **Sonny Lion**[1], **Albert Shih**[3], **Stéphane Aicardi**[3], **Philippe Zarka**[1,2], **Corentin Louis**[1,4], **Andrée Coffre**[2], **Laurent Lamy**[1,2], **Laurent Denis**[2], **Jean-Mathias Grießmeier**[5], **Jeremy Faden**[6], **Chris Piker**[6], **Nicolas André**[4], **Vincent Génot**[4], **Stéphane Erard**[1], **Joseph N Mafi**[7], **Todd A King**[7], **Jim Sky**[8], **Markus Demleitner**[9]

[1]LESIA, Observatoire de Paris, CNRS, PSL, Meudon, France. [2]Station de Radioastronomie de Nançay, Observatoire de Paris, CNRS, PSL, Université d'Orléans, Nançay, France. [3]DIO, Observatoire de Paris, CNRS, PSL, Paris, France. [4]IRAP, CNRS, Université Paul Sabatier, CNES, Toulouse, France. [5]LPC2E, CNRS, Université d'Orléans, Orléans, France. [6]Dep. Physics and Astronomy, University of Iowa, Iowa City, Iowa, USA. [7]IGPP, UCLA, Los Angeles, California, USA. [8]Radio Sky Publishing, USA. [9]Heidelberg Universität, Heidelberg, Germany.

Corresponding author: Baptiste Cecconi (baptiste.cecconi@observatoiredeparis.psl.eu)

MASER (Measurements, Analysis, and Simulation of Emission in the Radio range) is a comprehensive infrastructure dedicated to time-dependent low frequency radio astronomy (up to about 50 MHz). The main radio sources observed in this spectral range are the Sun, the magnetized planets (Earth, Jupiter, Saturn), and our Galaxy, which are observed either from ground or space. Ground observatories can capture high resolution data streams with a high sensitivity. Conversely, space-borne instruments can observe below the ionospheric cut-off (at about 10 MHz) and can be placed closer to the studied object. Several tools have been developed in the last decade for sharing space physics data. Data visualization tools developed by various institutes are available to share, display and analyse space physics time series and spectrograms. The MASER team has selected a sub-set of those tools and applied them to low frequency radio astronomy. MASER also includes a Python software library for reading raw data from agency archives.

## 1. Introduction

Low frequency radio data are providing remote proxies to remotely study energetic and unstable magnetised plasmas. In the solar system, all magnetized plasma environments are emitting radio emissions. The corresponding radio sources are non-thermal emission phenomena, and are not related to atomic and molecular transitions contrarily to electromagnetic emissions at higher frequencies. Their beaming pattern is also strongly anisotropic (Zarka 1998). The *low frequency* radio emissions are observed in the standard VLF (~3 kHz) to VHF (~30MHz) radio bands. The main radio sources of the solar system are the Sun, Jupiter and Saturn. The Earth, Uranus and Neptune are also hosting natural radio emissions. The planetary radio emissions are linked to the magnetospheric dynamics — i.e., auroral activity, radiation belts, etc. — as well as planetary atmospheres — i.e., lightning electromagnetic pulses.

The usual data product for low frequency radio emissions observations is a "dynamic spectrum" (a time varying spectrogram). Other products in use are high temporal resolution waveform snapshots (see, e.g., Briand et al 2016) and catalogues of events (with a radio bursts classification, see, e.g. Marques et al 2017). In this frequency range, it is not yet possible to build imaging radio telescopes, so that the main source of knowledge is this time-frequency representation of the data. Until recently each low frequency data provider was storing their data products in local formats, or using standard formats with local metadata dictionaries, which prevented interoperability. The NASA space physics community has promoted the Common Data Format (CDF, http://cdf.gsfc.nasa.gov) format with International Solar Terrestrial Program (ISTP, https://spdf.gsfc.nasa.gov/istp_guide/istp_guide.html) guidelines, for day to day usage and archiving. NASA's Planetary Data System (PDS) archive is now accepting CDF/ISTP as an archive format (King & Mafi 2018), and many space mission teams have adopted the same scheme. Ground based observatories are producing data collections reaching several TB per day (Lamy 2017). However, even with a common file format, downloading large data volumes for local processing is not optimal (e.g., long download delays) and should be avoided. There is thus a need for science-ready tools and standards that cover the needs of the low frequency radio astronomy community for time-dependent data.

## 2. The MASER collaboration

MASER (http://maser.lesia.obspm.fr) (Cecconi 2018) is a collaboration of teams throughout the world, whose aim is to facilitate the open access to science ready low frequency radio data. It is led by Observatoire de Paris (ObsParis) in France including people from LESIA (Laboratoire d'Etudes Spatiales et Instrumentation en Astrophysique) and PADC (Paris Astronomical Data Centre). It gathers scientists and software engineers from other space plasma and radio astronomy labs in France (Orléans, Nançay and Toulouse), and in the USA (University of Iowa, and University of California Los Angeles). Regular collaborations also exist with colleagues in Japan (Tohoku University) and Poland.

The MASER team at ObsParis is organized around 4 tasks: (a) data distribution; (b) codes and models; (c) infrastructure and interfaces; and (d) the *MaserLib* open-source repository. Task (a) is covering the full data lifecycle: from the production of the data (for ongoing and future data collection), the preparation of their distribution (data formatting, metadata, previews…), their validation (against the selected standards), the implementation of the access interfaces (web portal, virtual observatory, streaming interface…), as well as the curation of data when applicable. The details are described in a regularly updated Data Management Plan, following the open science policies developed at ObsParis (see Figure 1). The other French teams are working with ObsParis to implement similar policies on their data collections.

The other tasks deal with infrastructure and software development. Task (b) focusses on modelling codes (e.g., ray tracing code, radio observation modelling…), working with the science team to open the source code, share the simulation runs (through task (a)) and setup run-on-demand capabilities. Task (c) is the development and maintenance of the generic interoperable infrastructures in use to distribute the data collections managed in task (a) and (b). Task (d) is the development of the open source software libraries.

## 3. Data and Metadata Formats

MASER promotes the use of community standards for the data formats and metadata dictionaries. The space physics community is using CDF, whereas the Solar physics remote sensing community is using files formatted in Flexible Image Transport System (FITS, https://fits.gsfc.nasa.gov) (Pence 2010). Interoperability also requires enforcing adoption of standard metadata. MASER thus implements standards from the Heliophysics and Planetary Science communities: Space Physics Archive Search and Extracted (SPASE, http://spase-group.org) and ISTP metadata for Heliophysics; the Virtual European Solar and Planetary Access (VESPA, http://www.europlanet-vespa.eu) metadata for Planetary Sciences (Erard 2018).

## 4. Tools and Interfaces

The display tools and interfaces selected by MASER have been initially developed for space physics applications, as well as astronomy and solar system sciences. Space physics tools used by MASER have been developed by the University of Iowa: *Autoplot* (http://autoplot.org) (Faden 2010) and *Das2* (https://das2.org) (Piker 2018). We also use more generic technologies, such as VESPA, which provides a search interface that allows the discovery of data of interest for scientific users, and is based on International Virtual Observatory Alliance (IVOA) astronomy standards.

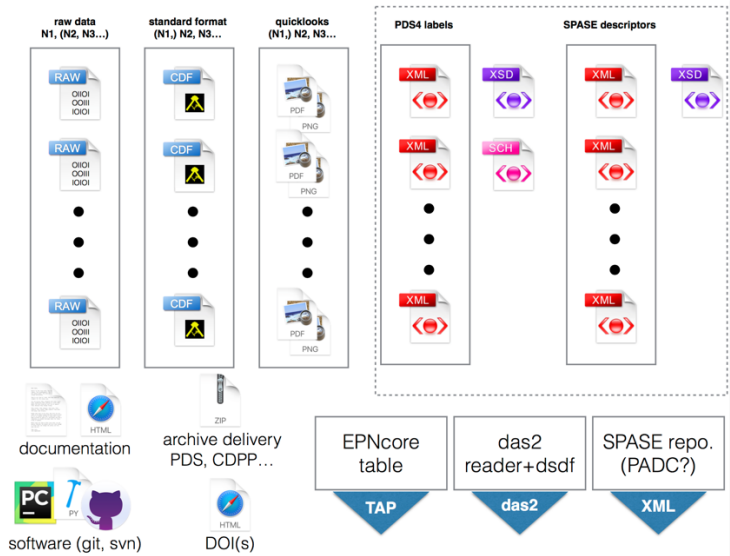

*Figure 1*. Synthetic description of the MASER data management plan.

### *4.1 Data streaming interface*

The driving issue of MASER is the *science ready* and *remote access* to low frequency radio astronomy data collections, and more specifically to long time-series or high-resolution datasets. For instance, each Solar transit observation (8 hours) data from the NewRoutine receiver of the Nancay Decameter Array (NDA) is stored in a 768 MB file, with 57,600 consecutive spectra (1 spectrum every 500 ms) (Lamy 2017). Displaying data on a typical computer screen requires about 2,000 pixels on the horizontal (i.e., temporal) axis, so that displaying the aforementioned Solar transit only requires transferring about 27 MB after temporal resampling (a reduction factor of about 30). As presented by Lamy (2017), some datasets from Nançay are reaching a few milliseconds of temporal resolution, so that the daily file size reaches several TBytes, and the data collection spans over several years. Remote visualisation of such data with adaptive temporal resolution streaming capabilities is thus needed.

The *Das2* data streaming technology allows to visualize data with a server-side time-axis resampling. The transmitted data are adjusted to the client temporal resolution, leading to a reduction of the data transfer over Internet. This reduces the delay for displaying the data, proportionally to the resampling rate, as specified by *das2* clients, such as *Autoplot* or the *das2py* library (https://github.com/das-developers/das2py). The installation and configuration of the *das2* server framework (https://github.com/das-developers/das2-pyserver), is simple and straightforward. Data providers have to develop a data reader script, which writes out the data as a *das2stream* into the local standard output for a given input time interval. The *das2stream* format is documented in its Interface Control Document (ICD) (Piker et al 2017).
In addition to the original das2 servers at University of Iowa, two other das2 servers are running to serve LESIA (http://voparis-das-maser.obspm.fr/das2/server) and Nançay (https://das2server.obs-nancay.fr/das2/server) datasets. Data readers are using the *maser4py* library (see section 5) for reading the data files from the local repositories. Figure 2 shows a dynamic spectrum of calibrated Cassini/RPWS/HFR data during the Jupiter flyby on December 31st 2000 and January 1st 2001, using Autoplot and accessing the data through the MASER/LESIA *das2* server.

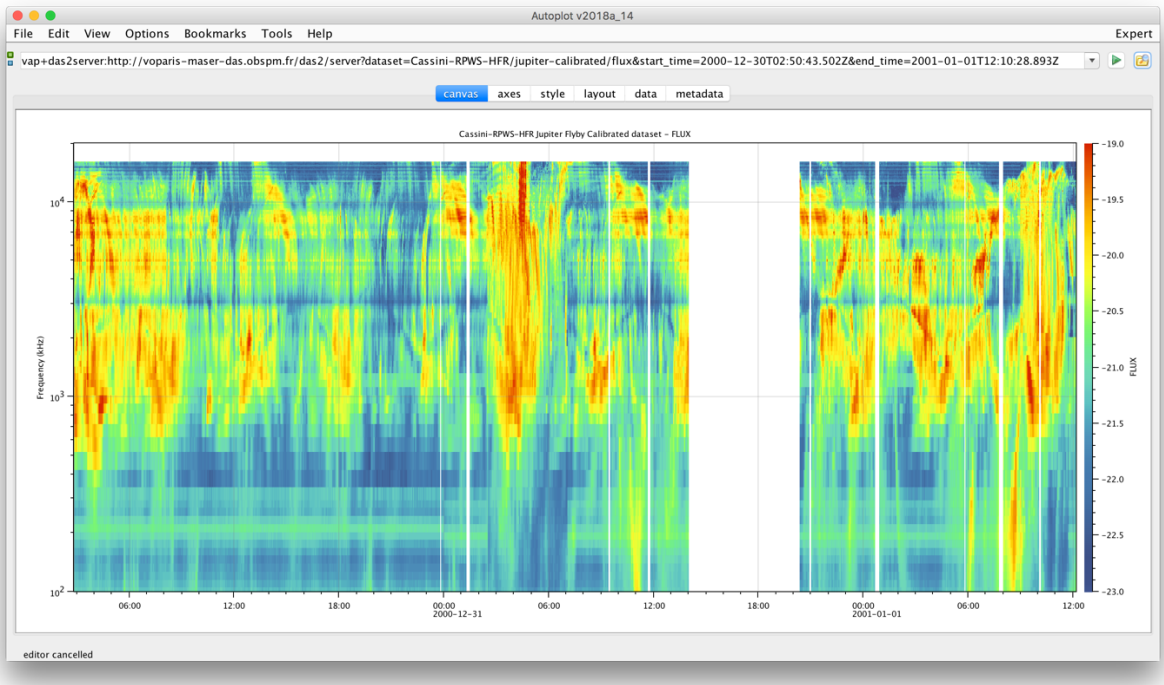

*Figure 2*. Time-Frequency spectrogram of radio emissions observed by Cassini/RPWS/HFR during the Jupiter flyby on December 31st 2000 and January 1st 2001, accessed through Autoplot and the MASER das2server interface.

### 4.2 Data discovery interface

VESPA is providing a data discovery framework with a metadata dictionary, a query protocol and a registry of services. Each VESPA service consists in a metadata table following the

EPNcore metadata dictionary (Erard 2018). Each row contains the metadata corresponding to a single product, including a data access URL. The VESPA services are running over the Table Access Protocol (TAP, http://www.ivoa.net/documents/TAP/) from the IVOA.

MASER teams are currently sharing data files (Raw, CDF or FITS formats) through VESPA. The VESPA services are built upon the Data Centre Helper Suite (DaCHS, http://dachs-doc.readthedocs.io) framework, and the tables are fed directly from reading the CDF or FITS headers. The VESPA main query portal (http://vespa.obspm.fr) also includes capabilities to interact directly with *Autoplot* (with the Simple Application Messaging Protocol (SAMP, http://www.ivoa.net/documents/SAMP/) of IVOA).

Several TAP servers dedicated to distributing MASER VESPA catalogue tables are available (http://voparis-tap-maser.obspm.fr at ObsParis, http://vogate.obs-nancay.fr in Nançay). They serve data collections from space mission with radio instruments (Cassini, Voyager, STEREO), from ground instruments (NDA) or modelled data (ExPRES, see section 6).

### 4.3 *Run-on-demand interface*

The IVOA has developed a computing job management system called Universal Worker Service (UWS, http://www.ivoa.net/documents/UWS/). MASER has implemented an instance of the Observatoire de Paris UWS System (OPUS, https://uws-server.readthedocs.io/en/latest/) available at https://voparis-uws-maser.obspm.fr. This server allows to submit jobs on a local computing cluster, either for automated data production pipelines (through a command line scripting interface), or for external users (through the web interface).

## 5. Maser4py Library

The *maser4py* library (https://github.com/maserlib/maser4py/) is providing data reader modules (for Python 3.6 and up) for legacy and non-standard format radio data collections. It currently includes modules for data collections hosted or produced by LESIA (Cassini/RPWS, Voyager/PRA, Solar Orbiter/RPW), by the Centre de Données de la Physique des Plasmas (CDPP, http://cdpp.eu) (Demeter, Interball, Viking (Swedish auroral mission), ISEE3, Wind), by the Planetary Plasma Interaction (PPI) node of NASA/PDS (Cassini/RPWS, Voyager/PRA), by the Nançay radio telescopes (NDA, NenuFAR), as well as by the radio amateur RadioJOVE project. It also includes generic modules developed for the ground segment of Solar Orbiter/RPW and a query interface for the HELIO-HFC (Heliophysics Integrated Observatory Feature Catalog) (Bonnin 2013). The *maser4py* library is open-source (GPLv3 license).

## 6. Modelling

The *Exoplanetary and Planetary Radio Emission Simulator* (ExPRES) code computes the geometric visibility of modelled auroral planetary radio source (Louis 2017, 2019). ExPRES is based on the Cyclotron Maser Instability (CMI) theory. It needs a planetary magnetic field model as well as parameters of the particle distributions in the modelled radio source. The code outputs time-frequency arrays of the visible auroral planetary radio source parameters (3D locus in the selected planetary frame and other radio source parameters). It is now used routinely to produce modelled daily spectrograms of simulated radio emissions induced by the Jovian Galilean satellites, for various observatory locations (Juno, Earth, STEREO). Precomputed simulation runs are available are available through different interfaces as defined in the MASER Data Management Plan (see Louis (2019) for more details).

ExPRES is open-source and its code is available at: https://github.com/maserlib/ExPRES. Run-on-demand is also available from the MASER OPUS server (see section 2.3). This computing interface requires an ExPRES JSON input configuration file. Examples of such configuration files are available through the web directory listing or virtual observatory catalogue: each of the precomputed file is provided with its input configuration file. The JSON input files must comply with the ExPRES JSON-schema specification, the current version of which is available at https://voparis-ns.obspm.fr/maser/expres/v1.0/schema#. Figure 3 shows the run-on-demand web interface where the user can manage his jobs. Figure 4 shows a simulation run compared to Juno/Waves data (courtesy of C. Louis).

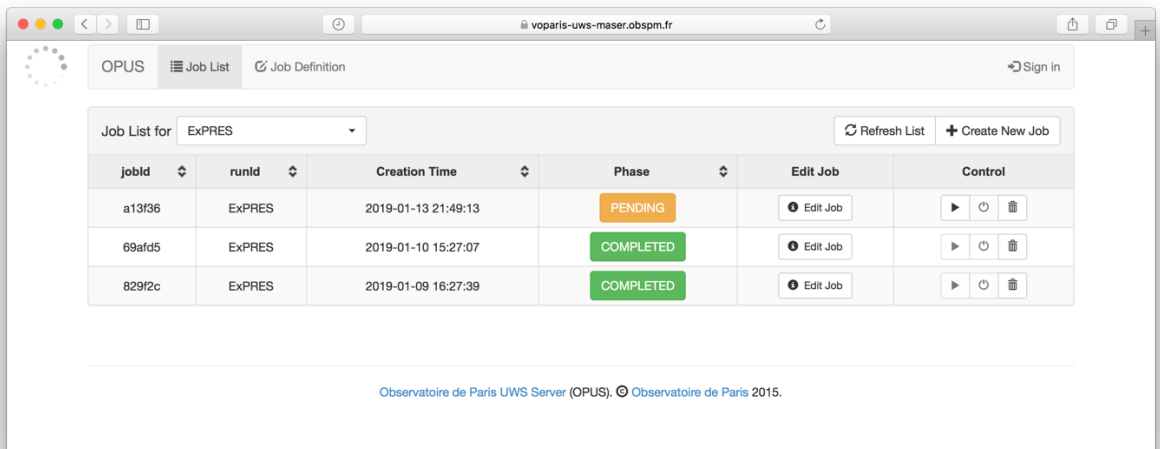

*Figure 3*. *MASER public run-on-demand interface. A few ExPRES runs are shown here. The user can manage his own jobs.*

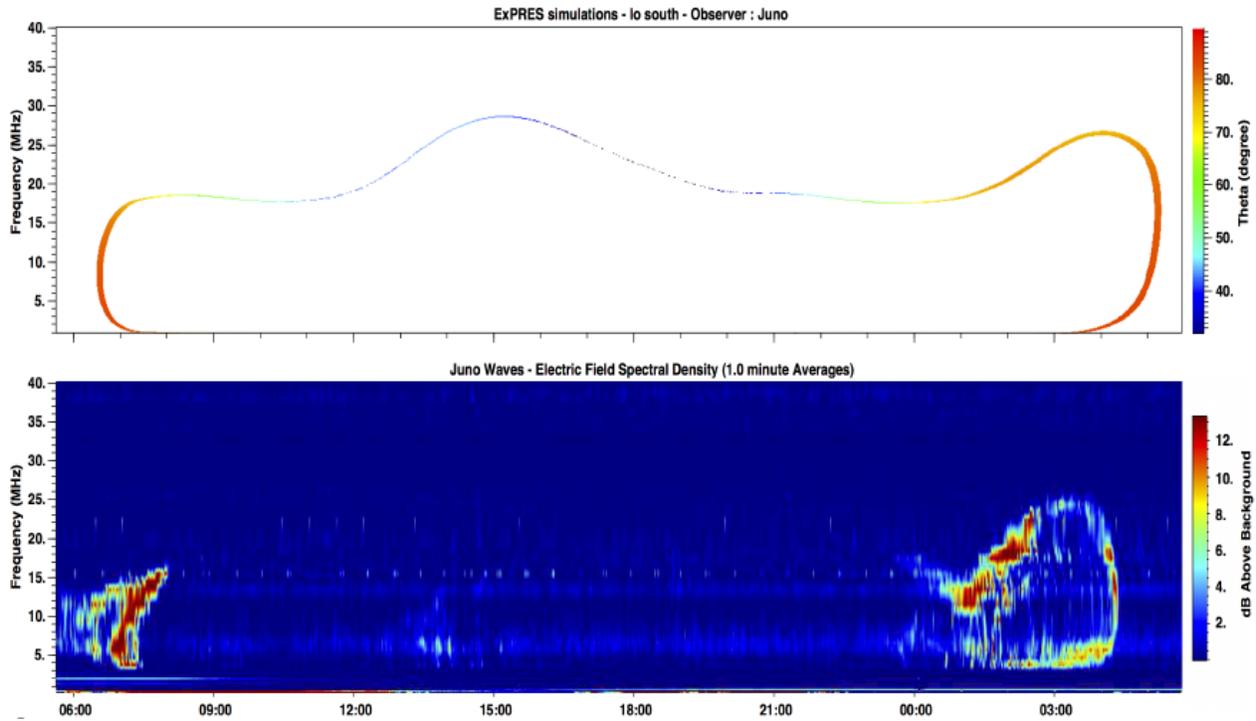

*Figure 4. Comparison of ExPRES simulation (top) and Juno/Waves data (bottom). Time-frequency spectrograms of Jovian radio emissions controlled by Io.*

We also plan to distribute the electromagnetic ray tracing code ARTEMIS-P (Gautier 2013) through MASER and UWS.

## 7. Applications

The usage of *das2* server interface with *Autoplot* improves data analysis and processing for low frequency radio astronomy. Within MASER, a few examples can be cited.

- The refurbishment of Voyager/PRA data (Cecconi 2018) has been consolidated by the *das2* server/*Autoplot* setup, allowing efficient and fast data browsing at all temporal scales;
- Distribution of low frequency data sets together with space observations (Lamy 2017);
- The NenuFAR (Zarka 2012) instrument is in commission phase in Nançay, and the team is testing VESPA as an internal data catalogue and *das2* server for fast data access;
- Juno-Ground-Radio (Cecconi 2016) is aggregating ground-based radio data from several observatories (France, USA, Ukraine, Japan, Poland…) and provides data supporting the Juno science team. The data files are distributed through VESPA, using CDF files when possible. *Das2* server interfaces are under study for collaborators in Ukraine and Poland.

## 8. Future Steps

New data readers will be continuously included in the *maser4py* library. The MASER team will also reach out to the community for participation. This requires a consolidation of the *maser4py* interfaces (classes and methods) and tests.

The ExPRES simulations are now used by the Juno/Waves instrument team. Discussions are ongoing with ESA, for using ExPRES as an observation planning support tool for the JUICE mission.

The need for a radio ground support has also been identified by the Solar Orbiter and Parker Solar Probe teams. The MASER tools and data collections are already available and serve those needs.

Finally, there is a growing need for community coordinated open source library and software developments (especially for python-based developments). Several groups are pushing for this, and MASER will participate to these efforts (e.g., http://openplanetary.co for planetary sciences; or the PyHC working group, http://heliopython.org).

In addition to the VESPA access and the *das2* server interface, we will follow the International Heliophysics Data Environment Alliance (https://ihdea.net) recommendation to implement HAPI (Heliophysics API) interfaces (Vandegriff 2018).

## Acknowledgments

The Europlanet H2020 Research Infrastructure project has received funding from the European Union's Horizon 2020 research and innovation programme under grant agreement No 654208. The teams also received support from Observatoire de Paris, CNES and CNRS/INSU through ASOV.